\documentclass[12pt]{article}
\usepackage{amsfonts,amsthm,amsmath}
\usepackage{siunitx}
\usepackage{url}
\usepackage{lscape}
\newtheorem{thm}{Theorem}
\newtheorem{prop}[thm]{Proposition}

\theoremstyle{remark}

\newcommand{\FF}{\mathbb{F}}

\DeclareMathOperator{\Aut}{Aut}
\DeclareMathOperator{\wt}{wt}
\begin{document}
\title{Classification of ternary  maximal self-orthogonal codes of length $25$}

\author{
Makoto Araya\thanks{Department of Computer Science,
Shizuoka University,
Hamamatsu 432--8011, Japan.
email: {\tt araya@inf.shizuoka.ac.jp}}
and
Masaaki Harada\thanks{
Research Center for Pure and Applied Mathematics,
Graduate School of Information Sciences,
Tohoku University, Sendai 980--8579, Japan.
email: \texttt{mharada@tohoku.ac.jp}.}
}

\maketitle

\begin{abstract}
Ternary  maximal self-orthogonal codes have been classified for lengths up to $24$.
In this note, we provide a complete classification of ternary maximal self-orthogonal 
codes of length $25$.
\end{abstract}

%%%%%%%%%%%%%%%  Section 1  %%%%%%%%%%%%%%%%%%
\section{Introduction}
\label{Sec:1}

Let $\FF_3 = \{0, 1, 2\}$ denote the finite field of order 3.
A \emph{ternary $[n,k]$ code} $C$ is a $k$-dimensional vector subspace
of $\FF_3^n$.
%All codes in this note are ternary.
The parameters $n$ and $k$ are called the \emph{length} 
and the \emph{dimension} of $C$, respectively.
The \emph{weight} $\wt(x)$ of a vector $x \in \FF_3^n$ is
the number of non-zero components of $x$.
A vector in $C$ is called a \emph{codeword} of $C$.
The minimum non-zero weight of all codewords in $C$ is called
the \emph{minimum weight} of $C$.
A  ternary $[n,k,d]$ code is a ternary $[n,k]$ code with minimum weight $d$.

The \emph{dual code} $C^{\perp}$ of a  ternary code
$C$ of length $n$ is defined as:
\[
C^{\perp}=
\{x \in \FF_3^n \mid x \cdot y = 0 \text{ for all } y \in C\},
\]
where $x \cdot y$ denotes the standard inner product of $x$ and $y$.
A  ternary code $C$ is \emph{self-orthogonal} if $C\subset C^\perp$, and 
$C$ is \emph{self-dual} if $C = C^\perp$.
A  ternary self-dual code of length $n$ exists if and only if
$n \equiv 0 \pmod 4$ with $n >0$.
A  ternary self-orthogonal code $C$ is \emph{maximal} if $C$ is
the only self-orthogonal code containing $C$.
A self-dual code is automatically maximal.
%The dimension of a maximal self-orthogonal code of
%length $n$ is a constant depending only on $n$.
%More precisely, 
A maximal self-orthogonal code of length $n$
has dimension $(n-1)/2$ if $n$ is odd and
$n/2-1$ if $n \equiv 2 \pmod 4$ (see~\cite{MPS}).

Two ternary codes $C$ and $C'$ are \emph{equivalent} 
if there is a monomial matrix $P$ such that
$C' = C \cdot P$, where $C \cdot P = \{ x P\:|\: x \in C\}$.
%and \emph{inequivalent} otherwise.
%We denote two equivalent codes $C$ and $D$ by $C \cong D$.
Ternary maximal self-orthogonal codes were classified in~\cite{MPS} for lengths up to $12$.
This classification was extended to lengths $13$, $14$, $15$ and $16$ in~\cite{CPS},
and lengths $17$, $18$, $19$ and $20$ in~\cite{PSW}
(see~\cite{HM-w} for lengths $18$ and $19$).
Subsequently, ternary self-dual codes of length 24 were classified in~\cite{HM-24}.
Building on this classification, ternary maximal self-orthogonal codes of lengths 
$21$, $22$ and $23$
were classified in~\cite{AHS}.

%In this note, 
%as another consequence of a classification of ternary self-dual codes of length $24$,
%we give a classification of ternary maximal  self-orthogonal $[25,12,9]$ codes.

In this note, we establish the following theorem, which is another 
consequence of the classification of ternary self-dual codes of length $24$.

\begin{thm}\label{thm}
There are $139613$ inequivalent  ternary maximal self-orthogonal 
$[25,12]$ codes.
Of these, $26$, $118984$ and $20603$ have minimum weights $9$, $6$ and $3$, respectively.
%There are $26, 118984$ and  $20603$ inequivalent  ternary maximal self-orthogonal 
%$[25,12,d]$ codes for $d=9,6$ and $3$, respectively.
\end{thm}

%Note that $9$ is the largest minimum weight among all ternary 
%maximal self-orthogonal $[25,12]$ codes.
This completes the classification of ternary maximal self-orthogonal codes for all 
lengths up to $25$.
To summarize, Table~\ref{Tab:C} lists the number $N(n)$ of 
inequivalent ternary maximal self-orthogonal codes of length $n$
together with the corresponding references for $3 \le n \le 25$
(see Proposition~\ref{prop:26-29} for lower bounds on 
$N(26)$, $N(27)$, $N(28)$, $N(29)$ and $N(30)$).

%%%%%%%%%%%%%%%%%%%%%%%%%%%%%%%
\begin{table}[thbp]
\caption{Ternary maximal self-orthogonal codes}
\label{Tab:C}
\centering
\medskip
{\small
%{\footnotesize
%{\scriptsize
\begin{tabular}{c|c|c||c|c|c}
\noalign{\hrule height1pt}
$n$ & $N(n)$ & Reference &
$n$ & $N(n)$ & Reference \\
\hline
 3 &      1 &\cite{MPS} &15 &     12 &\cite{CPS} \\
 4 &      1 &\cite{MPS} &16 &      7 &\cite{CPS} \\
 5 &      1 &\cite{MPS} &17 &     23 &\cite{PSW} \\
 6 &      2 &\cite{MPS} &18 &    160 &\cite{PSW} (see~\cite{HM-w}) \\
 7 &      1 &\cite{MPS} &19 &     56 &\cite{PSW} (see~\cite{HM-w}) \\
 8 &      1 &\cite{MPS} &20 &     24 &\cite{PSW} \\
 9 &      2 &\cite{MPS} &21 &    216 &\cite{AHS} \\
10 &      5 &\cite{MPS} &22 &  13625 &\cite{AHS} \\
11 &      3 &\cite{MPS} &23 &   2005 &\cite{AHS} \\
12 &      3 &\cite{MPS} &24 &    338 &\cite{HM-24} \\
13 &      7 &\cite{CPS} &25 & 139613 & Theorem~\ref{thm} \\
14 &     22 &\cite{CPS} & & & \\
\noalign{\hrule height1pt}
\end{tabular}
}
\end{table}
%%%%%%%%%%%%%%%%%%%%%%%%%%%%%%%

%%%%%%%%%%%%%%%%%%%%%%%%%%%%%%%%
\section{Classification methods}
\label{Sec:method}

In this section, we present the methodology used to establish the classification
given in Theorem~\ref{thm}.

%%%%%%%%%%%%%%%%%%%
\subsection{Shortening and lengthening}\label{Sec:inverse}
%CR The \emph{covering radius} of a ternary $[n,k]$ code $C$
%CR is the smallest integer $R$ such that spheres of radius $R$
%CR around codewords of $C$ cover the space $\FF_3^n$.
%CR It is known that the covering radius is the same as the largest weight of all coset leaders of $C$.
A \emph{shortened code} $C'$ of a  ternary code $C$ is the set of all codewords
in $C$ which are $0$ in a fixed coordinate with that
coordinate deleted.
A shortened code $C'$ of a ternary self-orthogonal $[n,k,d]$ code $C$
% with $d \ge 2$
is a ternary self-orthogonal $[n-1,k,d]$ code if the deleted coordinate
is a zero coordinate, and a ternary self-orthogonal $[n-1,k-1,d']$
code with $d' \ge d$ 
%CR and covering radius $R\ge d-1$
otherwise (see e.g., \cite[Section~4.1]{Brouwer-Handbook}).

We consider the inverse operation of shortening, which is referred to as \emph{lengthening}.
A  ternary self-orthogonal $[n,k,d]$ code gives $n$ shortened codes
and at least $k$ codes among them are self-orthogonal $[n-1,k-1,d']$ codes
with $d' \ge d$.
%CR and covering radius $R\ge d-1$.
Hence,
by applying the inverse operation of shortening,
any ternary self-orthogonal $[n,k,d]$ code can be constructed from some
ternary self-orthogonal $[n-1,k-1,d']$ code with $d' \ge d$ 
%CR and covering radius $R\ge d-1$ 
as follows.
Let $C'$ be a ternary self-orthogonal $[n-1,k-1,d']$ code with $d' \ge d$. 
%CR and covering radius $R\ge d-1$.
Up to equivalence,
we may assume that $C'$ has
a generator matrix of the form $
\left(\begin{array}{cc}
I_{k-1} & A
\end{array}\right)$, where $I_{k-1}$ denotes the
identity matrix of order $k-1$.
Then, up to equivalence,
a ternary self-orthogonal  $[n,k,d]$ code constructed from $C'$
by applying the inverse operation of shortening,
has the following generator matrix:
\begin{equation}\label{eq:gm}
G(A,b)=
\left(\begin{array}{c|ccc|cccccccc}
1&0 & \cdots& 0 &b_1 &\cdots&b_{n-k}\\
\hline
0& &       &      & &  &\\
\vdots& &I_{k-1}& & &A &\\
0& &       &     & &  &
\end{array}\right),
\end{equation}
where $b=(b_1,b_2,\ldots,b_{n-k}) \in\FF_3^{n-k}$
satisfies the following condition:
\begin{itemize}
\item[(C1)]
$\wt(b) \ge d-1$, $\wt(b) \equiv 2 \pmod 3$ and $A b^T=\mathbf{0}$.
\end{itemize}
Here $\mathbf{0}$ denotes the zero vector.
%In addition, we have the following basic properties:
%\begin{enumerate}
%\item
%$\langle G(A,b)\rangle \cong \langle G(A,2b)\rangle$.
%\item
%$\langle G(A,b)\rangle \cong \langle G(A,b')\rangle$ for $b-b' \in C'$,
%\end{enumerate}
%where $\langle G(A,b)\rangle$ denotes the ternary code with generator matrix $G(A,b)$.
It is sufficient to consider the vectors $b \in\FF_3^{n-k}$
satisfying the following condition:
\begin{itemize}
\item[(C2)]
the first nonzero component of $b$ is $1$,
\end{itemize}
since ternary codes with generator matrices
$G(A,b)$ and $ G(A,2b)$ are equivalent.

%%%%%%%%%%%%%%%%%%%
\subsection{Mass formula}
Suppose that $k \ge 2$ if $n$ is even, $n \ge 2k$,
$\varepsilon=1$ if $n \equiv 0 \pmod 4$
and 
$\varepsilon=-1$ if $n \equiv 2 \pmod 4$.
The number $T(n,k)$ of all distinct ternary self-orthogonal $[n,k]$
codes is given by:
\allowdisplaybreaks
\begin{equation}\label{eq:mf}
T(n,k)=
\begin{cases}
\displaystyle
\frac{(3^{n-k}-\varepsilon3^{\frac{n}{2}-k}+\varepsilon3^{\frac{n}{2}}-1) \prod_{i=1}^{k-1}(3^{n-2i}-1)}
{\prod_{i=1}^{k}(3^i-1)}  &\text{if $n$ is even,}
\\
\displaystyle
\frac{\prod_{i=0}^{k-1}(3^{2(\frac{n-1}{2}-i)}-1)}
{\prod_{i=1}^{k}(3^i-1)} &\text{if $n$ is odd}
\end{cases}
\end{equation}
(see~\cite{P65}).
%\begin{equation}\label{eq:mf}
%\begin{cases}
%\displaystyle
%\frac{(3^{n-k}-\varepsilon3^{n/2-k}+\varepsilon3^{n/2}-1) \prod_{i=1}^{k-1}(3^{n-2i}-1)}
%{\prod_{i=1}^{k}(3^i-1)}  &\text{ if $n$ is even,}
%\\
%\displaystyle
%\frac{\prod_{i=0}^{k-1}(3^{2((n-1)/2-i)}-1)}
%{\prod_{i=1}^{k}(3^i-1)} &\text{ if $n$ is odd.}
%\end{cases}
%\end{equation}
%&\frac{(3^{n-k}-3^{n/2-k}+3^{n/2}-1) \prod_{i=1}^{k-1}(3^{n-2i}-1)}
%{\prod_{i=1}^{k}(3^i-1)}  \text{ if $n \equiv 0 \pmod 4$,}\notag
%\\
%&\frac{(3^{n-k}+3^{n/2-k}-3^{n/2}-1) \prod_{i=1}^{k-1}(3^{n-2i}-1)}
%{\prod_{i=1}^{k}(3^i-1)}  \text{ if $n \equiv 2 \pmod 4$,}
%\\
The \emph{automorphism group} $\Aut(C)$ of a ternary code $C$ is the group of all
monomial matrices $P$ with $C = C \cdot P$.
Let ${\boldsymbol{C}_{n,k}}$ be a set of 
all inequivalent ternary self-orthogonal $[n,k]$ codes.
It is trivial that
\[
T(n,k)=\sum_{C \in {\boldsymbol{C}_{n,k}}} \frac{2^n n!}{|\Aut(C)|},
\]
and this is called the \emph{mass formula} for  ternary self-orthogonal $[n,k]$ codes.

%%%%%%%%%%%%%%%%%%%
\subsection{Equivalence testing}

For equivalence testing and automorphism group 
computation of ternary codes,
we employ  the approach given in~\cite[Section~7.3.3]{KO} as follows.

Let $C$ be a ternary  $[n,k]$ code and let $C(i)$ be the set of codewords of
weight $i$ in $C$.
Take a subset $W(C)$ of $\{1,2,\ldots,n\}$ such that
\[
\left\langle x \,\middle|\,  x \in \mathop{\cup}_{i \in W(C)} C(i) \right\rangle =C.
%\text{ and }
% C(i) \ne \emptyset \text{ for } i  \in W(C).
\]
Define
the vertex-colored digraph $\Gamma_{W(C)}(C)$ with the following vertex set:
\[
\left(\mathop{\cup}_{i \in W(C)} C(i) \right)\cup (\{1,2,\dots,n\}\times (\FF_3 \setminus\{0\}))
\]
and the following arc set:
\begin{align*}
&
\left\{(c,(j,c_j)), ((j,c_j),c) \,\middle|\, c=(c_1,c_2,\ldots,c_{n}) \in 
\mathop{\cup}_{i \in W(C)} C(i),  1 \le j \le n, c_j\ne 0\right\}
\\&
\cup \{((j,y),(j, 2y))\mid 1 \le j \le n,\ y \in \FF_3 \setminus \{0\}\}.
\end{align*}
Let $C$ and $C'$ be two ternary  $[n,k,d]$ codes.
Suppose that $W(C) = W(C')$. Then $C$ and $C'$ are equivalent
if and only if $\Gamma_{W(C)}(C)$ and $\Gamma_{W(C')}(C')$  are isomorphic as vertex-colored digraphs.
The order of the automorphism group of $\Gamma_{W(C)}(C)$ is the same as 
$|\Aut(C)|$.
We use \textsc{nauty}~\cite{nauty} for isomorphism testing and automorphism group 
computation of vertex-colored digraphs.

In our calculations in Section~\ref{Sec:result},
we took $W(C)=\{9\}$, $\{12\}$ or $\{24\}$ in most cases.
However, in one particular case, it was necessary to take
$W(C)=\{9,24\}$ in order to carry out the computation of the automorphism group.
This can also be carried out using the functions provided in \textsc{Magma}~\cite{Magma}.

%%%%%%%%%%%%%%%%%%%%%%%%%%%%%%%%
\section{Classification results}
\label{Sec:result}

In this section, we present the classification results obtained by the methodology described 
in the previous section.
Computer calculations in this section were performed using
by programs written in \textsc{C/C++} with
\textsc{nauty}~\cite{nauty} and \textsc{NTL}~\cite{NTL}
as well as  programs written in \textsc{Magma}~\cite{Magma}.

%%%%%%%%%%%%%%%%%%%
\subsection{Ternary maximal self-orthogonal $[25,12]$ codes with dual distances $1$}

There are two inequivalent ternary (extremal) self-dual  $[24,12,9]$ codes~\cite{LPS}. 
%, namely, the extended quadratic residue code  and the Pless symmetry
%code of length $24$~\cite{LPS}. 
%We denote these codes by $QR_{24}$ and $P_{24}$, respectively.
There are $166$ inequivalent ternary self-dual $[24, 12, 6]$ codes and
there are $170$ inequivalent ternary self-dual $[24, 12, 3]$ codes~\cite{HM-24}.
Note that $9$ is the largest minimum weight among all ternary self-dual $[24,12]$ codes.
%According to the order given in~\cite{Data},
%we denote these codes by $C_{24,6,i}$ $(i=1,2,\ldots,166)$
%and $C_{24,3,i}$ $(i=1,2,\ldots,170)$, respectively.

The \emph{dual distance} of a ternary code $C$ is defined as
the minimum weight of $C^\perp$.
% and it is denoted by $d^\perp$.
It is trivial that any  ternary maximal  self-orthogonal $[25,12]$ code with dual distance $1$ is
equivalent to the ternary code constructed as:
\[
\{(x_1,x_2,\ldots,x_{24},0) \mid (x_1,x_2,\ldots,x_{24}) \in C\},
\]
where $C$ is a self-dual code of length $24$.
Then we have the following:

\begin{prop}\label{prop:dd1}
There are two, $166$ and $170$ inequivalent ternary maximal  self-orthogonal
$[25,12,d]$ codes with dual distances $1$ for $d=9$, $6$ and $3$, respectively. 
\end{prop}

%%%%%%%%%%%%%%%%%%%
\subsection{Ternary self-orthogonal $[24,11]$ codes}

To find all ternary maximal self-orthogonal $[25,12]$ codes, which
require checking for equivalence by the lengthening operation, 
a classification of ternary self-orthogonal $[24,11]$ codes is necessary.
Moreover, any ternary self-orthogonal $[24,11]$ code is contained in some
ternary self-dual $[24,12]$ code (see e.g., \cite{MPS}).
By enumerating all ternary self-orthogonal $[24,11]$ subcodes of all inequivalent
ternary self-dual $[24,12]$ codes, we obtained candidates,
which were then subjected to equivalence testing.
After the equivalence testing described in the previous section,
we completed the classification of ternary self-orthogonal $[24,11]$ codes, 
as summarized in the following proposition.

\begin{prop}\label{prop:24-11}
There are $1373$, $2745294$ and  $339492$ inequivalent  ternary self-orthogonal 
$[24,11,d]$ codes for $d=9$, $6$ and $3$, respectively.
\end{prop}

Let ${\boldsymbol{C}_{24,11}}$ denote the set of all inequivalent 
ternary self-orthogonal $[24,11]$ codes.
As a check, we verified the mass formula:
\begin{align*}
\sum_{C \in {\boldsymbol{C}_{24,11}}}\frac{2^{24}24!}{|\Aut(C)|}
&=12850554292569078425974899530137600000\\
&=T(24,11),
\end{align*}
(see~\eqref{eq:mf} for the value $T(24,11)$).
The mass formula confirms
that there are no other 
ternary self-orthogonal $[24,11]$ codes, and thus the classification is complete.

%%%%%%%%%%%%%%%%%%%
\subsection{Ternary maximal self-orthogonal $[25,12]$ codes}

By applying the lengthening operation to the ternary self-orthogonal $[24,11]$ codes
given in Proposition~\ref{prop:24-11}, we obtained candidates for ternary maximal
self-orthogonal $[25,12]$ codes, which were then subjected to equivalence testing.
We tested the equivalence of these codes together with those given in
Proposition~\ref{prop:dd1} using the method described in the previous section.
After performing the equivalence testing, we completed the classification of
ternary maximal self-orthogonal $[25,12]$ codes, as stated in Theorem~\ref{thm}.

In Tables~\ref{Tab:Aut9}, \ref{Tab:Aut6} and \ref{Tab:Aut3},
we list the orders $|\Aut|$ of automorphism groups of 
self-orthogonal $[25,12,d]$ codes, together with the number $N$ of codes having each order
for $d=9$, $6$ and $3$, respectively.

%%%%%%%%%%%%%%%%%%%%%%%%%%%%%%%%
%\begin{table}[thbp]
%\caption{Orders of automorphism groups of self-orthogonal $[25,12,9]$ codes}
%\label{Tab:Aut9}
%\centering
%\medskip
%{\small
%%{\footnotesize
%%{\scriptsize
%\begin{tabular}{c|c||c|c||c|c||c|c}
%\noalign{\hrule height1pt}
%$|\Aut|$& $N$ & $|\Aut|$& $N$ & $|\Aut|$& $N$ & $|\Aut|$& $N$ \\
%\hline
%     2& 10 &     6& 10 &   20&  1 & 24288&  1 \\
%      4&  1 &    12&  2 & 10560&  1 & & \\
%\noalign{\hrule height1pt}
%\end{tabular}
%}
%\end{table}
%%%%%%%%%%%%%%%%%%%%%%%%%%%%%%%%

%%%%%%%%%%%%%%%%%%%%%%%%%%%%%%%
\begin{table}[thbp]
\caption{Orders of automorphism groups of self-orthogonal $[25,12,9]$ codes}
\label{Tab:Aut9}
\centering
\medskip
{\small
%{\footnotesize
%{\scriptsize
\begin{tabular}{c|c||c|c||c|c||c|c}
\noalign{\hrule height1pt}
$|\Aut|$& $N$ & $|\Aut|$& $N$ & $|\Aut|$& $N$ & $|\Aut|$& $N$ \\
\hline
     ${2}$& 10 &     ${2}.{3}$ & 10 &   ${2}^{2}5$&  1 & ${2}^{5}{3}.{11}.{23}$&  1 \\
      ${2}^{2}$&  1 &   $ {2}^{2}{3}$&  2 & ${2}^{6}{3}.{5}.{11}$&  1 & & \\
\noalign{\hrule height1pt}
\end{tabular}
}
\end{table}
%%%%%%%%%%%%%%%%%%%%%%%%%%%%%%%

%%%%%%%%%%%%%%%%%%%%%%%%%%%%%%%%
%\begin{table}[thbp]
%\caption{Orders of automorphism groups of self-orthogonal $[25,12,6]$ codes}
%\label{Tab:Aut6}
%\centering
%\medskip
%%{\small
%{\footnotesize
%%{\scriptsize
%\begin{tabular}{c|c||c|c||c|c||c|c||c|c}
%\noalign{\hrule height1pt}
%$|\Aut|$& $N$ &
%$|\Aut|$& $N$ &
%$|\Aut|$& $N$ &
%$|\Aut|$&$N$ &
%$|\Aut|$& $N$ \\
%\hline
%2& 83526 &216& 21 &3456& 10 &20736& 6 &552960& 1 \\
%4& 22399 &256& 62 &3888& 2 &23040& 2 &622080& 2 \\
%6& 143 &288& 114 &4096& 1 &24192& 1 &746496& 2 \\
%8& 7372 &384& 23 &4320& 2 &25920& 1 &1244160& 1 \\
%12& 187 &432& 34 &4608& 6 &27648& 1 &1492992& 2 \\
%16& 2484 &512& 15 &5184& 14 &31104& 2 &1935360& 1 \\
%18& 2 &576& 76 &5760& 2 &34560& 6 &2073600& 1 \\
%24& 207 &648& 6 &6144& 3 &41472& 4 &2661120& 1 \\
%32& 944 &768& 18 &6912& 4 &46080& 1 &3041280& 1 \\
%36& 42 &864& 13 &8640& 2 &51840& 1 &4852224& 1 \\
%40& 1 &1024& 8 &9216& 5 &62208& 4 &6842880& 1 \\
%48& 158 &1152& 36 &10368& 10 &69120& 3 &16174080& 1 \\
%64& 384 &1296& 6 &11520& 3 &98304& 1 &16588800& 1 \\
%72& 116 &1536& 6 &12096& 1 &103680& 2 &45619200& 1 \\
%96& 84 &1728& 18 &13824& 3 &124416& 1 &177914880& 1 \\
%108& 3 &1920& 1 &15552& 1 &207360& 3 &2134978560& 1 \\
%128& 148 &2048& 4 &16384& 1 &373248& 1 &144521625600& 1 \\
%144& 117 &2304& 19 &17280& 2 &414720& 2 & & \\
%192& 38 &2592& 13 &18432& 1 &483840& 1 & & \\
%\noalign{\hrule height1pt}
%\end{tabular}
%}
%\end{table}
%%%%%%%%%%%%%%%%%%%%%%%%%%%%%%%%

%%%%%%%%%%%%%%%%%%%%%%%%%%%%%%%
\begin{table}[thbp]
\caption{Orders of automorphism groups of self-orthogonal $[25,12,6]$ codes}
\label{Tab:Aut6}
\centering
\medskip
{\small
%{\footnotesize
%{\scriptsize
\begin{tabular}{c|c||c|c||c|c||c|c||c|c}
\noalign{\hrule height1pt}
$|\Aut|$& $N$ &
$|\Aut|$& $N$ &
$|\Aut|$& $N$ &
$|\Aut|$&$N$ &
$|\Aut|$& $N$ \\
\hline
${2}$ & 83526 & ${2}^{3}{3}^{3}$ & 21 & ${2}^{7}{3}^{3}$ & 
10 & ${2}^{8}{3}^{4}$ & 6 & ${2}^{12}{3}^{3}{5}$ & 1 \\
${2}^{2}$ & 22399 & ${2}^{8}$ & 62 & ${2}^{4}{3}^{5}$ & 
2 & ${2}^{9}{3}^{2}{5}$ & 2 & ${2}^{9}{3}^{5}{5}$ & 2 \\
${2}.{3}$ & 143 & ${2}^{5}{3}^{2}$ & 114 & ${2}^{12}$ & 
1 & ${2}^{7}{3}^{3}{7}$ & 1 & ${2}^{10}{3}^{6}$ & 2 \\
${2}^{3}$ & 7372 & ${2}^{7}{3}$ & 23 & ${2}^{5}{3}^{3}{5}$ & 
2 & ${2}^{6}{3}^{4}{5}$ & 1 & ${2}^{10}{3}^{5}{5}$ & 1 \\
${2}^{2}{3}$ & 187 & ${2}^{4}{3}^{3}$ & 34 & ${2}^{9}{3}^{2}$ & 
6 & ${2}^{10}{3}^{3}$ & 1 & ${2}^{11}{3}^{6}$ & 2 \\
${2}^{4}$ & 2484 & ${2}^{9}$ & 15 & ${2}^{6}{3}^{4}$ & 
14 & ${2}^{7}{3}^{5}$ & 2 & ${2}^{11}{3}^{3}{5}.{7}$ & 1 \\
${2}.{3}^{2}$ & 2 & ${2}^{6}{3}^{2}$ & 76 & ${2}^{7}{3}^{2}{5}$ & 
2 & ${2}^{8}{3}^{3}{5}$ & 6 & ${2}^{10}{3}^{4}{5}^{2}$ & 1 \\
${2}^{3}{3}$ & 207 & ${2}^{3}{3}^{4}$ & 6 & ${2}^{11}{3}$ & 
3 & ${2}^{9}{3}^{4}$ & 4 & ${2}^{8}{3}^{3}{5}.{7}.{11}$ & 1 \\
${2}^{5}$ & 944 & ${2}^{8}{3}$ & 18 & ${2}^{8}{3}^{3}$ & 
4 & ${2}^{10}{3}^{2}{5}$ & 1 & ${2}^{11}{3}^{3}{5}.{11}$ & 1 \\
${2}^{2}{3}^{2}$ & 42 & ${2}^{5}{3}^{3}$ & 13 & ${2}^{6}{3}^{3}{5}$ & 
2 & ${2}^{7}{3}^{4}{5}$ & 1 & ${2}^{9}{3}^{6}{13}$ & 1 \\
${2}^{3}{5}$ & 1 & ${2}^{10}$ & 8 & ${2}^{10}{3}^{2}$ & 
5 & ${2}^{8}{3}^{5}$ & 4 & ${2}^{9}{3}^{5}{5}.{11}$ & 1 \\
${2}^{4}{3}$ & 158 & ${2}^{7}{3}^{2}$ & 36 & ${2}^{7}{3}^{4}$ & 
10 & ${2}^{9}{3}^{3}{5}$ & 3 & ${2}^{10}{3}^{5}{5}.{13}$ & 1 \\
${2}^{6}$ & 384 & ${2}^{4}{3}^{4}$ & 6 & ${2}^{8}{3}^{2}{5}$ & 
3 & ${2}^{15}{3}$ & 1 & ${2}^{13}{3}^{4}{5}^{2}$ & 1 \\
${2}^{3}{3}^{2}$ & 116 & ${2}^{9}{3}$ & 6 & ${2}^{6}{3}^{3}{7}$ & 
1 & ${2}^{8}{3}^{4}{5}$ & 2 & ${2}^{11}{3}^{4}{5}^{2}{11}$ & 1 \\
${2}^{5}{3}$ & 84 & ${2}^{6}{3}^{3}$ & 18 & ${2}^{9}{3}^{3}$ & 
3 & ${2}^{9}{3}^{5}$ & 1 & ${2}^{10}{3}^{5}{5}.{11}.{13}$ & 1 \\
${2}^{2}{3}^{3}$ & 3 & ${2}^{7}{3}.{5}$ & 1 & ${2}^{6}{3}^{5}$ & 
1 & ${2}^{9}{3}^{4}{5}$ & 3 & ${2}^{12}{3}^{6}{5}.{11}.{13}$ & 
1 \\
${2}^{7}$ & 148 & ${2}^{11}$ & 4 & ${2}^{14}$ & 1 & ${2}^{9}{3}^{6}$ & 
1 & ${2}^{16}{3}^{6}{5}^{2}{11}^{2}$ & 1 \\
${2}^{4}{3}^{2}$ & 117 & ${2}^{8}{3}^{2}$ & 19 & ${2}^{7}{3}^{3}{5}$ & 
2 & ${2}^{10}{3}^{4}{5}$ & 2 & &\\
${2}^{6}{3}$ & 38 & ${2}^{5}{3}^{4}$ & 13 & ${2}^{11}{3}^{2}$ & 
1 & ${2}^{9}{3}^{3}{5}.{7}$ & 1 & &\\
\noalign{\hrule height1pt}
\end{tabular}
}
\end{table}
As a check, we verified the mass formula:
\begin{align*}
\sum_{C \in {\boldsymbol{C}_{25,12}}}\frac{2^{25}25!}{|\Aut(C)|}
&=25701205307660304745058529866383360000 \\
&=T(25,12),
\end{align*}
where ${\boldsymbol{C}_{25,12}}$
denotes the set of all inequivalent 
ternary maximal self-orthogonal $[25,12]$ codes (see Tables~\ref{Tab:Aut9}, \ref{Tab:Aut6} and \ref{Tab:Aut3}
for the orders $|\Aut(C)|$ and see~\eqref{eq:mf} for the value $T(25,12)$).
The mass formula confirms
that there are no other 
ternary maximal self-orthogonal $[25,12]$ codes, and thus the classification 
is complete.
The ternary maximal self-orthogonal codes of length $25$
are available electronically from 
\url{https://www.math.is.tohoku.ac.jp/~mharada/T25/}.

%%%%%%%%%%%%%%%%%
\section{Remarks}
We derive a (not necessarily tight) lower bound on the number of inequivalent
ternary maximal self-orthogonal codes of larger lengths by using the mass formula.
This is a classical technique in the study of self-dual codes (see e.g., \cite[pp.~52--53]{CP}), 
and we adapt it to the setting of ternary maximal self-orthogonal codes.
Let $C$ be a ternary self-orthogonal $[n,k]$ code.
Since $\{I_n,2I_n\}$ is a subgroup of the automorphism group $\Aut(C)$ of $C$,
we have that $|\Aut(C)| \ge 2$.
Let ${\boldsymbol{C}_{n,k}}$ denote a set of 
all inequivalent ternary self-orthogonal $[n,k]$ codes.
Since the mass formula gives that
\[
T(n,k)
=\sum_{C \in {\boldsymbol{C}_{n,k}}} \frac{2^n n!}{|\Aut(C)|}
\le  |{\boldsymbol{C}_{n,k}}| 2^{n-1} n!,
\]
we have that
\[
|{\boldsymbol{C}_{n,k}}| \ge 
\left\lceil \frac{T(n,k)}{2^{n-1}n!} \right\rceil,
%\frac{T(n,k)}{2^{n-1}n!}.
\]
where $\lceil x \rceil$ denotes the smallest integer greater than or equal to $x$.
We computed the values:
\[
\left\lceil \frac{T(n,k)}{2^{n-1}n!} \right\rceil
=
\begin{cases}
757009213 &\text{if } (n,k)=(26,12), \\
56074757  &\text{if } (n,k)=(27, 13),\\
2002670  &\text{if } (n,k)=(28,14), \\
82575085630 &\text{if } (n,k)=(29, 14), \\
4936926278278054 &\text{if } (n,k)=(30, 14).
\end{cases}
\]
These values yield the lower bounds shown below.

\begin{prop}\label{prop:26-29}
Let $N(n)$ denote the number  of 
inequivalent ternary maximal self-orthogonal codes of length $n$.
Then
\[
\begin{array}{ll}
N(26) \ge 757009213, &
N(27) \ge 56074757, \\
N(28) \ge 2002670, &
N(29) \ge 82575085630\text{ and } \\
N(30) \ge 4936926278278054.
\end{array}
\]
\end{prop}

% \begin{prop}\label{prop:26-29}
% \begin{enumerate}
% \item
% There are at least $757009213$  inequivalent ternary
% maximal self-orthogonal $[26,12]$ codes.
% \item
% There are at least $56074757$  inequivalent ternary
% maximal self-orthogonal $[27,13]$ codes.
% \item
% There are at least $2002670$  inequivalent ternary self-dual $[28,14]$ codes.
% \item
% There are at least $82575085630$  inequivalent ternary
% maximal self-orthogonal $[29,14]$ codes.
% \item
% There are at least $4936926278278054$  inequivalent ternary
% maximal self-orthogonal $[30,14]$ codes.
% \end{enumerate}
% \end{prop}

Finally, we turn to an application of 
a classification of ternary maximal self-orthogonal codes of length $25$.
By applying the method given in~\cite{HM-w} to this classification, 
it is a worthwhile project to complete a classification of 
weighing matrices of order $25$ and weight $9$.

%%%%%%%%%%%%%%%%
\bigskip
\noindent
\textbf{Acknowledgments.} 
This work is supported by JSPS KAKENHI Grant Numbers 23H01087 and 25K07111.
In this research work, we used the supercomputer of ACCMS, Kyoto University.

%%%%%%%%%%%%%%%%%%%  references  %%%%%%%%%%%%%%%%%%%%%%

\begin{landscape}
%%%%%%%%%%%%%%%%%%%%%%%%%%%%%%%
\begin{table}[thbp]
\caption{Orders of automorphism groups of self-orthogonal $[25,12,3]$ codes}
\label{Tab:Aut3}
\centering
\medskip
{\small
%{\footnotesize
%{\scriptsize
$
\begin{tabular}{c|c||c|c||c|c||c|c||c|c||c|c||c|c||c|c}
\noalign{\hrule height1pt}
$|\Aut|$& $N$ &
$|\Aut|$& $N$ &
$|\Aut|$& $N$ &
$|\Aut|$&$N$ &
$|\Aut|$& $N$ &
$|\Aut|$&$N$ &
$|\Aut|$& $N$ &
$|\Aut|$& $N$ \\
\hline
${2}^{2}{3}$&6716&${2}^{6}{3}^{3}$&217&${2}^{6}{3}^{4}{5}$&1&${2}^{8}{3}^{6}$&19&${2}^{9}{3}^{7}$&6&${2}^{14}{3}^{4}{5}$&1&${2}^{16}{3}^{6}$&1&${2}^{13}{3}^{10}$&1\\
${2}^{3}{3}$&4819&${2}^{8}{3}^{2}$&111&${2}^{10}{3}^{3}$&35&${2}^{10}{3}^{3}{7}$&1&${2}^{10}{3}^{5}{5}$&3&${2}^{10}{3}^{8}$&5&${2}^{13}{3}^{8}$&4&${2}^{14}{3}^{8}{5}$&1\\
${2}^{2}{3}^{2}$&10&${2}^{5}{3}^{4}$&55&${2}^{7}{3}^{5}$&38&${2}^{9}{3}^{4}{5}$&5&${2}^{14}{3}^{4}$&2&${2}^{9}{3}^{5}{5}.{11}$&3&${2}^{11}{3}^{7}{13}$&1&${2}^{13}{3}^{5}{5}^{2}{11}$&1\\
${2}^{4}{3}$&2302&${2}^{6}{3}^{2}{5}$&2&${2}^{9}{3}^{2}{7}$&1&${2}^{13}{3}^{3}$&5&${2}^{11}{3}^{6}$&15&${2}^{12}{3}^{5}{7}$&1&${2}^{14}{3}^{6}{5}$&1&${2}^{16}{3}^{6}{13}$&1\\
${2}^{2}{3}.{5}$&1&${2}^{10}{3}$&20&${2}^{8}{3}^{3}{5}$&6&${2}^{10}{3}^{5}$&18&${2}^{12}{3}^{4}{5}$&1&${2}^{11}{3}^{6}{5}$&3&${2}^{18}{3}^{5}$&1&${2}^{15}{3}^{9}$&2\\
${2}^{3}{3}^{2}$&505&${2}^{7}{3}^{3}$&188&${2}^{12}{3}^{2}$&5&${2}^{11}{3}^{3}{5}$&3&${2}^{16}{3}^{3}$&1&${2}^{12}{3}^{7}$&4&${2}^{11}{3}^{8}{5}$&1&${2}^{13}{3}^{8}{13}$&1\\
${2}^{5}{3}$&1055&${2}^{4}{3}^{5}$&1&${2}^{9}{3}^{4}$&59&${2}^{7}{3}^{7}$&1&${2}^{9}{3}^{6}{5}$&4&${2}^{11}{3}^{4}{5}.{11}$&1&${2}^{15}{3}^{7}$&2&${2}^{18}{3}^{6}{5}$&1\\
${2}^{3}{3}.{5}$&4&${2}^{9}{3}^{2}$&48&${2}^{6}{3}^{6}$&3&${2}^{9}{3}^{4}{7}$&1&${2}^{13}{3}^{5}$&5&${2}^{14}{3}^{4}{7}$&2&${2}^{13}{3}^{6}{13}$&1&${2}^{16}{3}^{9}$&1\\
${2}^{2}{3}.{11}$&1&${2}^{6}{3}^{4}$&85&${2}^{14}{3}$&1&${2}^{8}{3}^{5}{5}$&5&${2}^{10}{3}^{7}$&14&${2}^{17}{3}^{4}$&1&${2}^{12}{3}^{9}$&1&${2}^{11}{3}^{5}{5}^{2}{11}^{2}$&1\\
${2}^{4}{3}^{2}$&896&${2}^{7}{3}^{2}{5}$&1&${2}^{7}{3}^{4}{5}$&1&${2}^{12}{3}^{4}$&10&${2}^{12}{3}^{4}{7}$&1&${2}^{10}{3}^{7}{5}$&2&${2}^{13}{3}^{7}{5}$&1&${2}^{13}{3}^{6}{5}^{2}{11}$&1\\
${2}^{6}{3}$&490&${2}^{11}{3}$&10&${2}^{11}{3}^{3}$&12&${2}^{13}{3}^{2}{5}$&1&${2}^{11}{3}^{5}{5}$&4&${2}^{11}{3}^{5}{5}^{2}$&1&${2}^{16}{3}^{5}{7}$&1&${2}^{18}{3}^{8}$&1\\
${2}^{3}{3}^{3}$&12&${2}^{8}{3}^{3}$&121&${2}^{8}{3}^{5}$&45&${2}^{9}{3}^{6}$&17&${2}^{15}{3}^{4}$&3&${2}^{11}{3}^{8}$&3&${2}^{12}{3}^{8}{5}$&2&${2}^{14}{3}^{7}{5}.{11}$&2\\
${2}^{4}{3}.{5}$&1&${2}^{5}{3}^{5}$&14&${2}^{9}{3}^{3}{5}$&4&${2}^{11}{3}^{3}{7}$&1&${2}^{12}{3}^{6}$&8&${2}^{9}{3}^{7}{13}$&1&${2}^{11}{3}^{5}{5}^{2}{11}$&1&${2}^{20}{3}^{7}$&1\\
${2}^{5}{3}^{2}$&597&${2}^{10}{3}^{2}$&21&${2}^{13}{3}^{2}$&2&${2}^{10}{3}^{4}{5}$&3&${2}^{13}{3}^{4}{5}$&2&${2}^{12}{3}^{6}{5}$&2&${2}^{17}{3}^{5}{5}$&1&${2}^{17}{3}^{6}{5}.{11}$&1\\
${2}^{7}{3}$&202&${2}^{7}{3}^{4}$&81&${2}^{10}{3}^{4}$&36&${2}^{14}{3}^{3}$&1&${2}^{9}{3}^{8}$&2&${2}^{13}{3}^{7}$&2&${2}^{13}{3}^{9}$&3&${2}^{19}{3}^{7}{5}$&1\\
${2}^{4}{3}^{3}$&136&${2}^{12}{3}$&2&${2}^{7}{3}^{6}$&10&${2}^{11}{3}^{5}$&11&${2}^{10}{3}^{6}{5}$&6&${2}^{12}{3}^{4}{5}.{11}$&1&${2}^{12}{3}^{6}{5}.{11}$&2&${2}^{20}{3}^{8}$&1\\
${2}^{5}{3}.{5}$&1&${2}^{9}{3}^{3}$&54&${2}^{8}{3}^{4}{5}$&3&${2}^{12}{3}^{3}{5}$&1&${2}^{14}{3}^{5}$&1&${2}^{10}{3}^{9}$&1&${2}^{11}{3}^{8}{13}$&1&${2}^{18}{3}^{7}{13}$&1\\
${2}^{6}{3}^{2}$&391&${2}^{6}{3}^{5}$&25&${2}^{12}{3}^{3}$&10&${2}^{8}{3}^{7}$&4&${2}^{11}{3}^{7}$&7&${2}^{9}{3}^{6}{5}.{11}$&1&${2}^{14}{3}^{7}{5}$&2&${2}^{17}{3}^{10}$&1\\
${2}^{8}{3}$&91&${2}^{8}{3}^{2}{7}$&1&${2}^{9}{3}^{5}$&42&${2}^{9}{3}^{5}{5}$&2&${2}^{9}{3}^{6}{13}$&1&${2}^{15}{3}^{6}$&3&${2}^{15}{3}^{5}{5}^{2}$&1&${2}^{18}{3}^{10}$&1\\
${2}^{5}{3}^{3}$&242&${2}^{11}{3}^{2}$&13&${2}^{10}{3}^{3}{5}$&2&${2}^{13}{3}^{4}$&2&${2}^{12}{3}^{5}{5}$&4&${2}^{12}{3}^{8}$&2&${2}^{15}{3}^{8}$&1&${2}^{16}{3}^{8}{5}.{11}$&2\\
${2}^{7}{3}^{2}$&233&${2}^{8}{3}^{4}$&77&${2}^{8}{3}^{4}{7}$&1&${2}^{10}{3}^{6}$&14&${2}^{16}{3}^{4}$&1&${2}^{13}{3}^{6}{5}$&2&${2}^{14}{3}^{5}{5}.{11}$&1&${2}^{22}{3}^{9}$&1\\
${2}^{4}{3}^{4}$&17&${2}^{9}{3}^{2}{5}$&1&${2}^{14}{3}^{2}$&1&${2}^{12}{3}^{3}{7}$&1&${2}^{9}{3}^{7}{5}$&1&${2}^{17}{3}^{5}$&1&${2}^{11}{3}^{7}{5}.{11}$&1&${2}^{21}{3}^{7}{5}.{11}$&1\\
${2}^{3}{3}.{5}.{11}$&1&${2}^{5}{3}^{6}$&1&${2}^{7}{3}^{5}{5}$&3&${2}^{9}{3}^{5}{7}$&1&${2}^{13}{3}^{6}$&4&${2}^{14}{3}^{7}$&2&${2}^{13}{3}^{8}{5}$&1&${2}^{24}{3}^{9}$&1\\
${2}^{5}{3}^{2}{5}$&1&${2}^{7}{3}^{3}{7}$&1&${2}^{11}{3}^{4}$&17&${2}^{16}{3}.{5}$&1&${2}^{10}{3}^{5}{5}^{2}$&2&${2}^{15}{3}^{5}{5}$&1&${2}^{17}{3}^{7}$&1&${2}^{29}{3}^{8}{5}$&1\\
${2}^{9}{3}$&39&${2}^{13}{3}$&2&${2}^{12}{3}^{2}{5}$&1&${2}^{12}{3}^{5}$&13&${2}^{11}{3}^{5}{13}$&1&${2}^{12}{3}^{7}{5}$&2&${2}^{21}{3}^{3}{7}$&1&&\\
\noalign{\hrule height1pt}
\end{tabular}$
}
\end{table}
%%%%%%%%%%%%%%%%%%%%%%%%%%%%%%%
\end{landscape}

\end{document}